# PatchWorkPlot: simultaneous visualization of local alignments across multiple sequences

Maria Pospelova[1] & Yana Safonova[1,2,*]

[1] Computer Science and Engineering Department, Pennsylvania State University, State College, PA 16802, USA
[2] Huck Institutes of Life Science, Pennsylvania State University, State College, PA 16802, USA
[*] Corresponding author: yana@psu.edu

## Abstract

### Motivation

Revealing structural variations across sequences of closely related individuals or species is crucial for understanding their diversification mechanisms and roles.

### Results

We developed PatchWorkPlot, a tool for visualization of pairwise alignments of multiple annotated sequences as dot plots combined into a single matrix.

### Availability and implementation

PatchWorkPlot is implemented using Python 3 and is publicly available at GitHub: github.com/yana-safonova/PatchWorkPlot.

## Introduction

Development of sequencing technologies and genome assembly methods have enabled reconstruction of previously inaccessible and repetitive genomic regions[1–3]. Analysis of structural variations in such sequences across closely related subjects or species is an important step in understanding their role and diversification mechanisms. A dot plot representing all local alignments between two sequences remains the main tool for initial visual assessment of sequence similarities and structures[4,5]. A dot plot of two sequences $S_1$ and $S_2$ is a collection of line segments corresponding to all local alignments. A line segment representing an alignment of substrings of lengths $l_1$ and $l_2$ starts at $(x, y)$ and extends to $(x+l_1, y+l_2)$, where $x$ and $y$ are starting alignment positions in $S_1$ and $S_2$, respectively.

Availability of genomes from closely related species and population-wide data for the same species raises a question about a convenient way to combine results of local alignments of more than two sequences. While linear genome diagrams[6,7] might be a tool of choice for non-repetitive sequences, it is less informative for highly repetitive sequences. To overcome this challenge, we present PatchWorkPlot, a tool that analyzes multiple sequences, computes their pairwise alignments using sensitive local alignment tools, and visualizes the results in a matrix format for intuitive assessment. PatchworkPlot produces high-quality ready-to-publish figures and has a convenient interface for fine-tuning the results.

## Methods

### Software description

PatchWorkPlot includes two main steps: computing alignments and visualizing them as dot plots. Alignment is performed using an existing tool, and the current version supports LastZ[8] and YASS[9]. Once alignments are computed, PatchWorkPlot considers the orientation of the first sequence as the main one and orients other sequences with respect to it. If the longest alignments to the first sequence are in reverse orientation, the corresponding sequence will be reversed. PatchWorkPlot then generates dot plots for all sequence pairs and incorporates annotations if provided. All pairwise alignments are saved, allowing the tool to reuse them when rerun on the same output directory, significantly reducing the total running time.

### Input & output files

PatchWorkPlot takes a configuration file in the .CSV format as an input. Each input sequence is described as a line in the configuration file. The configuration file has three mandatory columns (`Fasta`: paths to sequences in the FASTA format, `SampleID`: unique sequence IDs, `Label`: sequence labels that will be shown in the final plot) and two optional columns (`Annotation`: paths to annotations in the BED format, `Strand`: orientations of the input sequences). If orientations are provided, Patchwork plot will use them; otherwise, it will determine them automatically as described above. PatchWorkPlot also works with a simplified input where all sequences are combined into a single FASTA, unique identifiers are shown as headers, and all annotations are combined into a single BED file. PatchWorkPlot reports pairwise dot plots combined into a triangular matrix as well as individual pairwise dot plots as PNG and PDF files.

### Visualization routine

PatchWorkPlot visualizes pairwise dot plots and arranges them as an upper (lower) triangular matrix, where a cell ($i$, $j$) shows a dot plot corresponding to alignments of $i$-th and $j$-th input sequences if $i \leq j$ ($i \geq j$). Dot plots on the main diagonal correspond to self-alignments of the input sequences. Each alignment passing the minimum length threshold (`--min-len`, default: 5 kbp) is shown as a colored line segment. PatchWorkPlot determines the alignment color based on its percent identity (PI), provided it falls within the range set by the `--min-pi` and `--max-pi` parameters (default: 85% and 100%). If the PI value is outside this range, the closest boundary value is used instead. It then projects the PI value into a Python colormap[10] (default: `rainbow`) to assign the color. A different Python colormap or a single color can be specified via '`--cmap`' or '`--color`' options, respectively. A user can also change the orientation of the colormap ('`--reverse`', default: True) and the width of alignment lines ('`--lwidth`', default: 1), or make the .PNG file transparent ('`--transparent`').

If paths to BED files were provided as an input, the user can specify the '`--show-annot`' option to add annotations for each sequence. Annotations are visualized on the side of the matrix as rectangles spanning the corresponding start and end positions, and an annotation unit is colored in black until a color is specified in the BED file.

**Fig. 1** shows examples of PatchWorkPlot outputs. The lower triangular plot on the left shows alignments of five immunoglobulin (IG) heavy chain loci of feline species (two haplotypes of the puma, the clouded leopard, the bobcat, the domestic cat) with lengths ranging from 1.2 to 2.3 Mbp, and the upper triangular plot on the right shows alignments of five IG heavy chain loci of great apes (the bonobo, the human, the gorilla, the Bornean orangutan, the Sumatran orangutan) with lengths 1.0-1.4 Mbp. Both plots show positions of IG genes computed using IgDetective[11] in black, the right plot also shows positions and Alu and LINE/L1 repeats computed using RepeatMasker[12] in green and red, respectively. Both PatchWorkPlot examples illustrate how the tool allows one to visually assess the structure of segmental duplications within and across the species and compare them with positions of important genetic elements.

## Results

PatchWorkPlot is designed to visualize complex genomic regions with a high density of structural variations. The tool allows the user to fine tune visualization parameters and add annotations enabling visual assessment of correspondence between annotated genetic elements and structural variations. PatchWorkPlot is implemented in Python 3 and requires little installation efforts.

PatchWorkPlot performance depends on the performance of the chosen alignment tool and the complexity of input sequences. On average, the visualization step takes 1% of the total running time. E.g., the running time on highly repetitive feline IG heavy chain loci (**Fig. 1**; 15 pairwise alignments) was 45 min 43 sec including alignment using LastZ and and 30 sec using previously computed alignments (8-core CPU, 16 Gb RAM). The running time on less repetitive great ape IG heavy chain loci (**Fig. 1**; 15 pairwise alignments) was 8 min 49 sec (6 sec) with (without) LastZ alignment on the same machine. The running time and memory usage increase quadratically with the number and length of sequences.

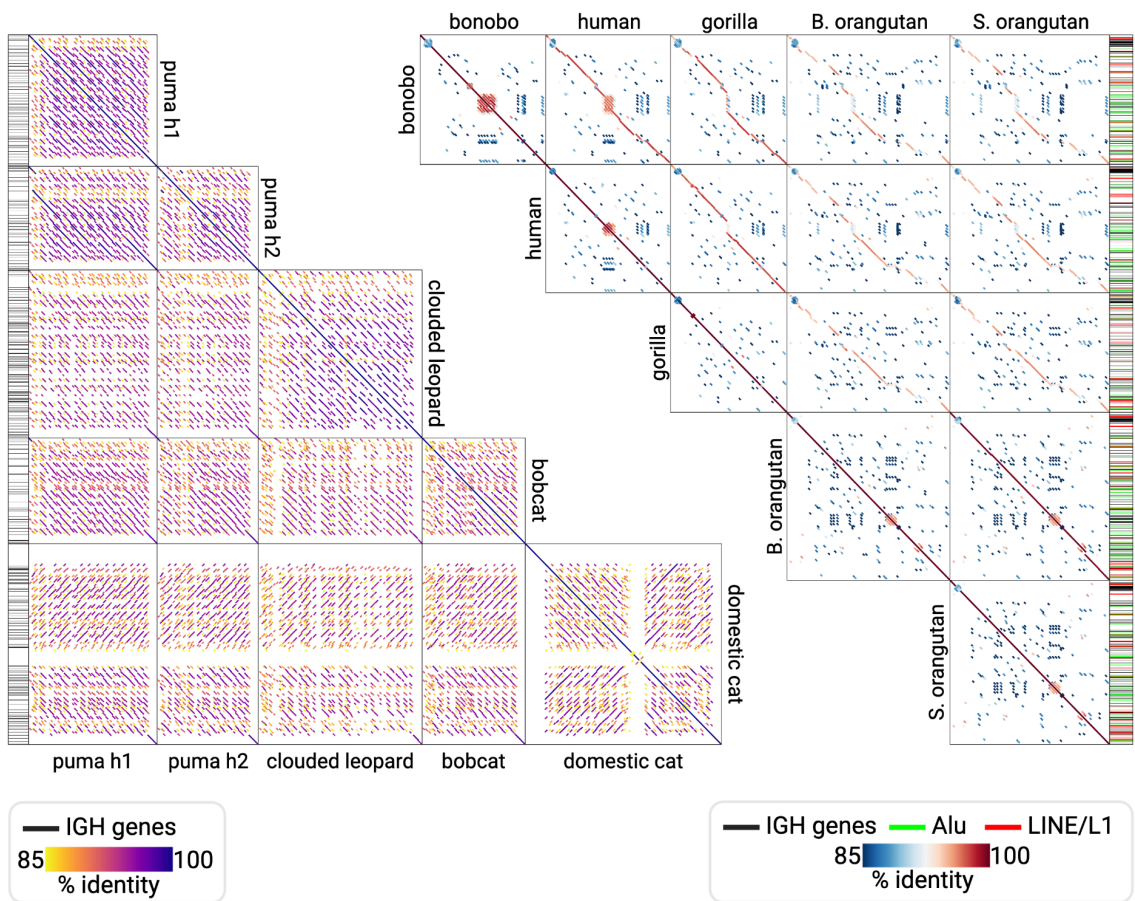

**Figure 1. Illustration of PatchWorkPlot work.** The left plot shows a lower triangular matrix representing alignments of five immunoglobulin heavy chain loci of four feline species (the puma haplotype 1: GCA_028749985.3; the puma haplotype 2: GCA_028749965.3, the clouded leopard: GCA_028018385.1; the bobcat: GCF_022079265.1; the domestic cat: GCA_013340865.1). The right plot shows an upper triangular matrix representing alignments of five immunoglobulin heavy chain loci of great ape species (the bonobo: GCA_029289425.3; the human: T2T-CHM13; the gorilla: GCA_029281585.3; the Bornean orangutan: GCA_028885625.3; the Sumatran orangutan: GCA_028885655.3). Annotations are shown as bars attached to the left or right side of the matrix. Plots were generated by the tool without manual modifications, command line parameters on the top and legends on the bottom of each plot were added manually for illustration purposes.